\documentclass[twocolumn,aps,prl,superscriptaddress]{revtex4}
\makeatletter

\newcommand{\Rmnum}[1]{\expandafter\@slowromancap\romannumeral #1@}
\makeatother
\usepackage{amsmath}
\usepackage{graphicx}
\usepackage{subfigure}
\usepackage{color}
\usepackage{amsfonts}
\usepackage{tikz}
\usepackage{esint}
\usepackage{mathrsfs}
\usepackage{hyperref}
\hypersetup{
     colorlinks = true,
     citecolor  = blue,  
     linkcolor  = red 
}
\begin{document}

\title{Phase diagrams of tunable spin-orbit coupled Bose-Einstein condensates}
\author{Zhi Lin}
\email{zhilin18@ahu.edu.cn}
\affiliation{School of Physics and Materials Science, Anhui University, Hefei 230601, P. R. China}
\affiliation{Department of Physics and State Key Laboratory of Surface Physics, Fudan University, Shanghai 200433, P.R. China}

\begin{abstract}
 We analytically calculate phase boundaries of tunable spin-orbit coupled  BECs with effective two-body interactions by using variational method.
 Phase diagrams for periodically driving $^{87}\rm{Rb}$ and  $^{23}\rm{Na}$ systems are presented, respectively, which display several characteristic features contrast with those of undriven systems. In the $^{87}\rm{Rb}$ BECs, the critical density $n_{c}$ (density at quantum tricritical point) can be dramatically reduced in some parameter regions, thus the prospect of observing this intriguing quantum tricritical point is greatly enlarged. Moreover, a series of quantum tricritical points emerge quasi-periodically with increasing the Raman coupling strength for fixed $^{87}\rm{Rb}$ density. In  the $^{23}\rm{Na}$ BECs, two hyperfine states of  $^{23}\rm{Na}$ atoms can be miscible due to driving under proper parameters. As a result, $^{23}\rm{Na}$ systems can stay in the stripe phase with small Raman frequency at typical density. This expands the region of stripe phase in the phase diagram. In addition,  there is no quantum tricritical point in such $^{23}\rm{Na}$ system , which is different from $^{87}\rm{Rb}$ system.
\end{abstract}

\maketitle
\section{Introduction\label{sec:1}}
Spin-orbit coupling(SOC) plays an important role in various areas of physics \cite{zhai-rev}, and engineering SOC with cold atoms enable us to study novel quantum phases of matter \cite{intrinsic-anoma-Hall,exotic-topo-orders,soc2df-Dirac-point1,soc2dfDirac-point2,soc2db} and novel transport currents \cite{Z2-type-semimetals,Weyl-semimetals}.  Depending on using atoms with internal or external degree of  freedom, SOC can be classified as internal SOC and external SOC \cite{esoc1,esoc}, respectively. These two types of SOC have been realized in  ultracold gas experiment. More specifically, by using hyperfine states (internal degree of freedom) of an alkali atom as pseudospins, one-dimensional internal SOC (equal Rashba and Dresselhaus SOC) for neutral bosonic atom \cite{sob} or fermionic atom \cite{sof1,sof2} has been realized for several years, and the two-dimensional internal SOC for ultracold $^{40}\rm{K}$  fermionic  gases \cite{soc2df-Dirac-point1,soc2dfDirac-point2} and bosonic gases $^{87}\rm{Rb}$ \cite{soc2db} has also been realized in recent years.  The external SOC  for ultracold   $^{23}\rm{Na}$ has also been realized in optical superlattice by choosing the two lowest eigenstates of the double-well potential as the pseudospin up and down component \cite{esoc}. The remarkable features of this scheme for realizing external SOC are that this scheme does not require near resonant light and can be applied to any atomic species \cite{esoc}.  Accompanying experimental advances, there has been significant theoretical progress in understanding the SOC of cold bosonic \cite{bose-so1,bose-so2,li-1} and fermionic \cite{fermi-so1,fermi-so2,fermi-so3,fermi-so4} gases.  The most fascinating prediction for spin-orbit coupled Bose-Einstein condensates (BECs) is the existence of a nontrivial stripe state \cite{bose-so1,bose-so2,li-1}. This state has been indirectly observed  \cite{obser-t} in internal spin-orbit coupled $^{87}\rm{Rb}$ BECs and directly observed in external spin-orbit coupled $^{23}\rm{Na}$ BECs \cite{obser-na}. Moreover, this intriguing stripe state has also been directly observed in $^{87}\rm{Rb}$ BECs (without SOC interaction) which are trapped at the intersection of two cavities.

On the other hand, Floquet engineering has become one of the hottest topics \cite{floquet1,floquet2,floquet3,floquet4} in cold atom systems, owing to its ability to engineer novel interactions. In driven optical lattice systems, Floquet technique has been utilized very successfully in various experiments with ultracold atoms. These experiments include  dynamic localization \cite{dynamic-l1,dynamic-l2,dynamic-l3}, controlling the superfluid-Mott insulator quantum phase transition \cite{C-sf-mi1,C-sf-mi2} of bosonic atoms, resonant coupling of Bloch bands \cite{Bloch-b1,Bloch-b2,Bloch-b3,Bloch-b4}. It has also been used to dynamically create  kinetic frustration\cite{Eckardt,frus2},
to realize artificial magnetic fields and topological band structures \cite{frus2,topol-1,topol-2,topol-3,topol-4,topol-5}. Besides, Floquet technique has also been used for tuning \cite{tunable-so} or inducing \cite{induce} spin orbit coupling  in quantum gases systems via  periodically modulated interaction strength.

Recently, I. B. Spielman's group has realized tunable internal SOC \cite{tunable-so} in $^{87}\rm{Rb}$  BECs that provides a way to connect these two important quantum control methods (using Raman laser to induce internal SOC and Floquet engineering) in ultracold atomic gases. Here, the tunable internal SOC is induced by periodically modulated the Raman coupling strength $\Omega(t)$.  The tunable internal SOC experiment stimulates us to study the phase diagrams of the driven $^{87}\rm{Rb}$ and $^{23}\rm{Na}$ BECs.  Although the two-body interactions cannot be ignored in a real experimental system, only the single-particle Hamiltonian is involved in effective Floquet Hamiltonian obtained by previous works \cite{tunable-so,zhang}.

In this paper,  we reveal that the effective Floquet Hamiltonian includes a novel effective two-body interaction by considering the two-body interactions in undriven Hamiltonian. And then, the phase diagrams of this effective Floquet Hamiltonian can be obtain via variational wave function. The characteristic features of phase diagrams of tunable spin-orbit coupled $^{87}\rm{Rb}$ and $^{23}\rm {Na}$ BECs are presented.
 In $^{87}\rm{Rb}$ BECs, the critical density $n_{c}$ will be reduced  dramatically and the value of critical density can be decreased at the typical experimental range of densities. By fixing $^{87}\rm{Rb}$ density,  quantum tricritical points appear quasi-periodically as increasing the value of dimensionless Raman coupling strength $\Omega_{R}/\omega$, owing to the fact that the SOC strength is a quasi-periodic function of $\Omega_{R}/\omega$.  It is qualitatively different from the undriven case that the two components of $^{23}\rm {Na}$ atoms are miscible in some parameter regions. In this miscible region, systems can stay in the stripe phase with small Raman frequency at the typical density  ($n\approx 10^{13} \rm{cm}^{-3}$) and the region of stripe phase is enhanced by increasing the value of density. Moreover, in contrast to $^{87}\rm{Rb}$ BECs, there is no quantum tricritical point in such spin-orbit coupled $^{23}\rm{Na}$ BECs.

The remainder of the paper is organized as follows. In section \ref{sec-2}, we will introduce the effective Floquet Hamiltonian. In section \ref{sec-3}, the mean-field approach for tunable spin-orbit coupled BECs are presented.  Section \ref{sec-4} is devoted to obtain the phase diagrams of $^{87}\rm{Rb}$ and $^{23}\rm{Na}$ BECs, respectively. The summary and concluding remarks are presented in  section \ref{sec-5}.
\section{Effective Floquet Hamiltonian\label{sec-2}}
In  I. B. Spielman's experiment \cite{tunable-so}, the frequency difference $\Delta \omega$ of two Raman lasers was set near Zeeman splitting frequency $\omega_{z}\approx 15 \rm{MHz}$,  and  $\delta_{0} = \Delta \omega -\omega_{Z}$ denotes the experimentally tunable detuning. Using the rotating wave approximation, the single-particle Hamiltonian including both the kinetic and Raman coupling contributions can be written as \cite{sob,tunable-so,rotating-wave}
\begin{equation}
\hat{H}_{0}=\left(\frac{\hbar^{2}q^{2}_{x}}{2m}+E_{L}\right)\hat{1}
+\frac{\hbar\Omega(t)}{2}\hat{\sigma}_{x}+\frac{\hbar\delta_{0}}{2}\hat{\sigma}_{z}-\alpha_{0}q_{x}\hat{\sigma}_{z}\label{H}
\end{equation}
where, $\Omega(t)=\Omega_{0}+\Omega_{R}\cos(\omega t)\approx E^{\ast}_{A}E_{B}$ is the Raman coupling strength, $E_{A,B}$ are the complex-valued optical electric field strengths, the driven frequency $\omega$ is of the order of \rm{kHz}, and $\hat{\sigma}_{x,y,z}$ are the Pauli
matrices. The recoil momentum, recoil energy, and SOC strength are denoted by  $k_{0}$, $E_{L}=\hbar^{2}k^{2}_{0}/2m$, and $\alpha_{0}=\hbar^{2}k_{0}/m$, respectively. In experiment, the recoil momentum $k_{0}$ are fixed by the momentum
transfer of the two Raman lasers. The effective Floquet Hamiltonian \cite{Eckardt,floquet2} can be obtained by using high frequency approximation, where the high frequency approximation is not in conflict with the rotating wave approximation, owing to the fact that the driven  frequency $\omega$ is about three orders of magnitude smaller than  the frequency difference $\Delta\omega$.

 After using the Floquet theory $\mathcal{H}_{F}=\hat{H}_{0}-i\hbar\partial_{t}$, the effective single-particle Hamiltonian satisfies the relation \cite{Eckardt}
\begin{equation}
\hat{h}_{0}=\langle \hat{U}^{\dag}(t)\mathcal{H}_{F}\hat{U}(t)\rangle_{T} \label{eff1}.
\end{equation}
where the unitary matrix $\hat{U}(t)$ reads
\begin{equation}
\hat{U}(t)=\exp{\left(-\frac{i}{\hbar}W(t)\hat{\sigma}_{x}\right)}\label{eff2}.
\end{equation}
Here the time-periodic Hermitian operator $\hat{W}(t)$ reads
\begin{equation}
 W(t)=\int^{t}_{0}\nu(\tau)d\tau -\langle \int^{t}_{0}\nu(\tau)d\tau\rangle_{T}\label{eff3},
\end{equation}
 $\langle A\rangle_{T}\equiv\int^{T}_{0}A(t) dt/T$ is the average in time space, and $\nu(t)=\hbar \Omega_{R}\cos(\omega t)/2$ is the time dependent part of Hamiltonian.
Using Eqs.~(\ref{eff1}), (\ref{eff2}), and (\ref{eff3}),  the effective single-particle Hamiltonian reads ($\hbar=m=1$)
\begin{equation}
\hat{h}_{0}=\left(\frac{q^{2}_{x}}{2}+\frac{k_{0}^{2}}{2}\right)\hat{1}
+\frac{\Omega_{0}}{2}\hat{\sigma}_{x}+\alpha\frac{\delta_{0}}{2}\hat{\sigma}_{z}-\alpha k_{0}q_{x}\hat{\sigma}_{z}, \label{h0}
\end{equation}
where $\alpha=J_{0}\left(\Omega_{R}/\omega\right)$ is the renormalized coefficient.

So far, the effective two-body interactions have not been addressed in this tunable spin-orbit coupled BECs. We will demonstrate the effective two-body interaction Hamiltonian by considering the two-body interactions in the Floquet Hamiltonian $\mathcal{H}_{F}$. To obtain the effective two-body interactions, we first need to introduce the form of two-body interaction for spin-1/2 Bose gases. The two-body interactions can be written as \cite{Li}
\begin{equation}
H_{\rm int}=\sum_{\sigma,\sigma^{\prime}}\frac{1}{2}\int d\mathbf{r} g_{\sigma,\sigma^{\prime}}\rho_{\sigma}(\mathbf{r})\rho_{\sigma^{\prime}}(\mathbf{r}),
\end{equation}
where the spin-dependent density operators are given by
\begin{equation}
\rho_{\pm}(\mathbf{r})=\frac{1}{2}\sum_{j}\left(1\pm\sigma_{z,j}\right)\delta(\mathbf{r}-\mathbf{r}_{j}),
\end{equation}
and  $j=1,\cdots, N$ is the particle index.
 Since $\hat{U}(t)$ is the unitary matrix, we can obtain the relation
\begin{equation}
\!\!\!\hat{U}^{\dag}\!(t)\rho_{\sigma}(\mathbf{r})\rho_{\sigma^{\prime}}(\mathbf{r})\hat{U}\!(t)\!
\!=\!\hat{U}^{\dag}\!(t)\rho_{\sigma}(\mathbf{r})\hat{U}\!(t)\hat{U}^{\dag}\!(t)\rho_{\sigma^{\prime}}(\mathbf{r})\hat{U}\!(t).
\end{equation}
We can also obtain
\begin{eqnarray}
\rho^{\prime}_{\pm}&=& \hat{U}^{\dag}(t)\rho_{\pm}(\mathbf{r})\hat{U}(t)=\frac{1}{2}\sum_{j}\left\{1\pm\left[\cos(2\chi)\sigma_{z,j}\right.\right.\nonumber\\
&&\left.\left.+\sin(2\chi)\sigma_{y,j}\right]\right\}\delta(\mathbf{r}-\mathbf{r}_{j})\qquad(\pm=\uparrow, \downarrow),
\end{eqnarray}
with $\chi=\Omega_{R}/\left(2\omega\right)\sin(\omega t)$. By taking the second quantization
$\sum_{j} A \rightarrow \int d\mathbf{r}_{j}\psi^{\dag}(\mathbf{r}_{j})A \psi(\mathbf{r}_{j}) $,
 the density operator can be written as
\begin{eqnarray}
\hat{\rho}^{\prime}_{\pm}&=&\int d\mathbf{r}_{j}\frac{1}{2}\psi^{\dag}(\mathbf{r}_{j})
\left[1\pm\left(\cos(2\chi)\sigma_{z,j} \right.\right.\nonumber\\
&&\qquad+\left.\left.\sin(2\chi)\sigma_{y,j}\right)\right]\psi(\mathbf{r}_{j})\delta(\mathbf{r}-\mathbf{r}_{j}),
\end{eqnarray}
where $\psi=\left(\psi_{\uparrow},\psi_{\downarrow}\right)^{T}$ is the two-component field operator.

In general, the time-dependent two-body interactions are given by \cite{Coleman}
\begin{equation}
\hat{H}_{\rm int}(t) =\frac{1}{2}\sum_{\sigma,\sigma^{\prime}}\int d\mathbf{r} g_{\sigma,\sigma^{\prime}}:\hat{\rho}_{\sigma}^{\prime}(\mathbf{r})\hat{\rho}^{\prime}_{\sigma^{\prime}}(\mathbf{r}):\label{time-two-body},
\end{equation}
where $:\ldots :$ is the normal order. In this general case,
by taking the time-average of Eq.~(\ref{time-two-body}),
the effective two-body interactions read
\begin{eqnarray}
\hat{H}^{\rm eff}_{\rm int}&=&\int d\mathbf{r}\left\{ \frac{g^{\ast}_{\uparrow,\uparrow}}{2}\psi^{\dag}_{\uparrow}\psi^{\dag}_{\uparrow}\psi_{\uparrow}\psi_{\uparrow} +
\frac{g^{\ast}_{\downarrow,\downarrow}}{2}\psi^{\dag}_{\downarrow}\psi^{\dag}_{\downarrow}\psi_{\downarrow}\psi_{\downarrow}\right.\nonumber \\
&&\qquad\qquad \qquad+g^{\ast}_{\uparrow,\downarrow}\psi^{\dag}_{\uparrow}\psi_{\uparrow}\psi^{\dag}_{\downarrow}\psi_{\downarrow}\nonumber \\
&&\left.\qquad-\frac{g^{\ast}}{2}\left(\psi^{\dag}_{\downarrow}\psi^{\dag}_{\downarrow}\psi_{\uparrow}\psi_{\uparrow}
+\psi^{\dag}_{\uparrow}\psi^{\dag}_{\uparrow}\psi_{\downarrow}\psi_{\downarrow}\right)\right\},\label{h1}
\end{eqnarray}
and the effective interaction strengths are given by
\begin{eqnarray}
g^{\ast}_{\uparrow,\uparrow}&=&\frac{g_{\uparrow,\uparrow}+g_{\downarrow,\downarrow}}{8}
\left[3+J_{0}\left(\frac{2\Omega_{R}}{\omega}\right)\right]\nonumber \\
&&\!+\frac{g_{\uparrow,\uparrow}\!-\!g_{\downarrow,\downarrow}}{2}\!J_{0}\!\left(\!\frac{\Omega_{R}}{\omega}\!\right)\!
\!+\!\frac{g_{\uparrow,\downarrow}}{4}\!\left[\!1\!-\!J_{0}\!\left(\!\frac{2\Omega_{R}}{\omega}\!\right)\!\right]\!,
\end{eqnarray}
\begin{eqnarray}
g^{\ast}_{\downarrow,\downarrow}&=&\frac{g_{\uparrow,\uparrow}+g_{\downarrow,\downarrow}}{8}
\left[3+J_{0}\left(\frac{2\Omega_{R}}{\omega}\right)\right]\nonumber \\
&&\!-\frac{g_{\uparrow,\uparrow}\!-\!g_{\downarrow,\downarrow}}{2}\!J_{0}\!\left(\!\frac{\Omega_{R}}{\omega}\!\right)\!
\!+\!\frac{g_{\uparrow,\downarrow}}{4}\!\left[\!1\!-\!J_{0}\!\left(\!\frac{2\Omega_{R}}{\omega}\!\right)\!\right]\!,
\end{eqnarray}

\begin{equation}
g^{\ast}_{\uparrow,\downarrow}\!\!=\!\!\frac{g_{\uparrow,\uparrow}\!+\!g_{\downarrow,\downarrow}}{4}
\!\left[\!1\!-\!J_{0}\!\left(\!\frac{2\Omega_{\!R}}{\omega}\!\right)\!\right]
\!+\frac{g_{\uparrow,\downarrow}}{2}\!\!\left[\!1\!+\!J_{0}\!\left(\!\frac{2\Omega_{\!R}}{\omega}\!\right)\!\right],
\end{equation}
\begin{equation}
g^{\ast}=\frac{g_{\uparrow,\uparrow}+g_{\downarrow,\downarrow}-2g_{\uparrow,\downarrow}}{8}
\left[1-J_{0}\left(\frac{2\Omega_{R}}{\omega}\right)\right].\label{g_eff}
\end{equation}
The new type of two-body interaction
$-(g^{\ast}/2) ( \!\psi^{\dag}_{\downarrow}\psi^{\dag}_{\downarrow}\psi_{\uparrow}\psi_{\uparrow}$
$+\psi^{\dag}_{\uparrow}\psi^{\dag}_{\uparrow}\psi_{\downarrow}\psi_{\downarrow}\! )$
appears in this effective two-body Hamiltonian.  At below, we will reveal that this novel term can enlarge the parameter region of stripe phase ( phase \Rmnum{1} ) for $^{23}$\rm{Na} BECs, but leads to the opposite consequence for $^{87}$\rm{Rb} BECs. Noteworthily, if we do not choose the normal order for two-body interactions, the Gross-Pitaevskii equation obtained from this effective Hamiltonian Eqs.~(\ref{h0},\ref{h1}) is the same as C. Zhang' work \cite{zhang}.

\section{ Phase diagram: mean-field approach \label{sec-3}}
In this section, we first formulate a mean-field energy function for the effective Floquet Hamiltonian, using reasonable wave function ansatz.
Then, the possible ground-state configuration  can be determined by minimizing the energy function of per particle for different cases.
\subsection{The mean-field energy function for per particle\label{sec-31}}
Using the Gross-Pitaevskii mean-field approach \cite{Li}, the energy function relevant to effective Floquet Hamiltonian [ Eq.~(\ref{h0}) and Eq.~(\ref{h1}) ] can be written as
\begin{eqnarray}
E&&\!=\!\int d\mathbf{r}\left[\left(
                  \begin{array}{cc}
                    \psi^{\ast}_{\uparrow} & \psi^{\ast}_{\downarrow} \\
                  \end{array}
                \right)\hat{h}_{0}\left(
                              \begin{array}{c}
                                \psi_{\uparrow} \\
                                \psi_{\downarrow} \\
                              \end{array}
                            \right)+\frac{g^{\ast}_{\uparrow,\uparrow}}{2}|\psi_{\uparrow}|^{4}
                           +\frac{g^{\ast}_{\downarrow,\downarrow}}{2}|\psi_{\downarrow}|^{4}\right. \nonumber\\
                             &&\left.+g^{\ast}_{\uparrow,\downarrow}|\psi_{\uparrow}|^{2}|\psi_{\downarrow}|^{2}
                            \!-\!\frac{g^{\ast}}{2}\left(\psi^{\ast}_{\downarrow}\psi^{\ast}_{\downarrow}\psi_{\uparrow}\psi_{\uparrow}
+\psi^{\ast}_{\uparrow}\psi^{\ast}_{\uparrow}\psi_{\downarrow}\psi_{\downarrow}\right)\right].\label{e-1}
\end{eqnarray}
If we label $g^{\ast}_{\uparrow,\uparrow}=g+\delta g$ and $g^{\ast}_{\downarrow,\downarrow}=g-\delta g$, then the total detuning energy is given by
\begin{equation}
E_{\rm d }=\int d\mathbf{r}\frac{1}{2}\left[\delta g \left(n_{\uparrow}+n_{\downarrow}\right)+\alpha\delta_{0}\right]
\left(n_{\uparrow}-n_{\downarrow}\right).
\end{equation}
We are only interested in the situation with $\delta = \delta g \overline{n} +\alpha\delta_{0}=0$ (take an approximation via replacing $n_{\uparrow}+n_{\downarrow}$ with $\overline{n}$) \cite{zhai-rev}.
This novel two-body interaction does not lead to detuning in systems. Therefore, the ansatz wave function \cite{li-1}
\begin{equation}
 \left(\!\begin{array}{c}
  \psi_{\uparrow}\\
  \psi_{\downarrow} \\
   \end{array}
    \!\right)\!\!=\!\sqrt{\!\frac{N}{V}\!}\!\left[\!C_{1}e^{ik_{1}x}\!\left(\!
                  \begin{array}{c}
                    \cos\theta \\
                    \!-\!\sin\theta \\
                  \end{array}
                \!\right)\!\!+\!C_{2}e^{\!-\!ik_{1}x}\!\left(\!
                  \begin{array}{c}
                    \sin\theta \\
                    \!-\!\cos\theta \\
                  \end{array}
                \!\right)\!\!\right]\!\label{ansatz }
\end{equation}
is still valid, where $N$ is the total number of atoms and $V$ is the volume of the system. For given density $n=N/V$, the
variational parameters are $C_{1}$, $C_{2}$, $k_{1}$ and $\theta$ ($2\theta\!=\!\arccos\left(k_1/\alpha k_{0}\right)$ and $ 0 \leq\theta \leq
\pi/4 $). Their values are determined by  minimizing energy in Eq.~(\ref{e-1}) with the normalization condition
$\sum_{i=\uparrow,\downarrow}\int d^{3}r|\psi|^{2}_{i}=N$ (choosing $|C_{1}|^{2}+|C_{2}|^2 =1$ ).
With the help of this ansatz wave function, the dimensionless per particle energy $\epsilon$ ( divided by $k^{2}_{0}$ ) reads
\begin{eqnarray}
\epsilon&=&\frac{1}{2}-\frac{\Omega^{\ast}_{0}}{2}\frac{\sqrt{\alpha^{2} k^{2}_{0}-k^{2}_{1}}}{|\alpha|
k_{0}}-F(\beta)\frac{k_{1}^{2}}{2\alpha^{2}k_{0}^{2}}\nonumber \\
&&+\left(G_{1}-G_{2}\xi\right)\left(1+2\beta\right),\label{e-3}
\end{eqnarray}
where we have defined two dimensionless parameters $\xi=[1-J_{0}\left(2\Omega_{R}/\omega\right)]/[1+3J_{0}\left(2\Omega_{R}/\omega\right)]$,
$\beta=|C_{1}|^{2}|C_{2}|^{2}$($\beta \in \left[0,1/4\right]$), and the function
\begin{equation}
F(\beta)=\left(\alpha^{2}-2G_{2}\left(1+\xi\right)\right)+
4\left(G_{1}+G_{2}\left(2-\xi\right)\right)\beta,
\end{equation}
with the dimensionless Raman frequency $\Omega^{\ast}_{0}=\Omega_{0}/k^{2}_{0}$, and the two dimensionless interaction parameters $G_{1}$ $=(n/8k^{2}_{0})(g^{\ast}_{\uparrow\uparrow}+g^{\ast}_{\downarrow\downarrow}+2g^{\ast}_{\uparrow\downarrow})$, $G_{2}=(n/8k^{2}_{0})(g^{\ast}_{\uparrow\uparrow}+g^{\ast}_{\downarrow\downarrow}-2g^{\ast}_{\uparrow\downarrow})$. It is easy to check that when switch off the modulation of the Raman coupling strength ($\Omega_{R}/\omega=0$ and $\xi=0$),  the dimensionless per particle energy $\epsilon$ in Eq.~(\ref{e-3}) is the same as the energy per particle of the undriven system \cite{li-1}.

\subsection{The possible ground-state configuration of tunable spin-orbit coupled BECs  \label{sec-32}}
Before considering the driven spin-orbit coupled BECs, we would like to give here a brief review of the ground state of undriven spin-orbit coupled BECs. In undriven systems, there are three possible phases, i.e., stripe phase ( phase \Rmnum{1}) with $k_{1}\neq 0$, $\beta=1/4$ and hence $\langle\sigma_{z}\rangle=0$, the separated phase (phase \Rmnum{2}) with $k_{1}\neq 0$, $\beta=0$, and hence $\langle\sigma_{z}\rangle\neq 0$, and zero momentum phase (phase \Rmnum{3}) with  $k_{1} =0$, $\beta=0$ and hence  $\langle\sigma_{z}\rangle=0 $\cite{li-1}.

In the section~\ref{sec-31}, the energy per particle $\epsilon$ as a function with the variational parameters $k_{1} =0$ and $\beta=0$  are obtained by using the wave function ansatz. Next, we will analyze the possible ground-state configuration of tunable spin-orbit coupled BECs by minimizing $\epsilon$. In order to make the discussion easier, we introduce two useful parameters  $a=n\left(g_{\uparrow,\uparrow}+g_{\downarrow,\downarrow}\right)/k^{2}_{0}$ ( $a>0$ for repulsive interaction) and $x=g_{\uparrow,\downarrow}/\left( g_{\uparrow,\uparrow}+g_{\downarrow,\downarrow}\right)$. And then, dimensionless parameters $G_{1}$, $G_{2}$, and the function $F(\beta)$ are rewritten as
\begin{eqnarray}
G_{1}&=& \frac{a}{32}\left[5+6x-\left(1-2x\right)J_{0}\left(\frac{2\Omega_{R}}{\omega}\right)\right], \\
G_{2}&=& \frac{a}{32}\left(1-2x\right)\left[1+3J_{0}\left(\frac{2\Omega_{R}}{\omega}\right)\right], \\
F(\beta)&\!=\!&\alpha^{2} +\frac{a}{8}\!\bigg[\!(6+4x)\beta-\!\left(1\!-\!2x\right)\!\times
 \nonumber \\
&&\left.\left(\!1+J_{0}\!\left(\!\frac{2\Omega_{R}}{\omega}\!\right)
-6J_{0}\!\left(\!\frac{2\Omega_{R}}{\omega}\!\right)\beta\right) \!\right]\!.
\end{eqnarray}
The derivative of $F(\beta)$ with respect to $\beta$ is given by
\begin{equation}
\!\!\!\partial_{\beta}F(\beta)\!=\!\frac{a}{8}\!\left[\!6+6J_{0}\!\left(\!\!\frac{2\Omega_{R}}{\omega}\!\!\right)\!
\!+\!4x\!\left(\!1\!-\!3J_{0}\!\left(\!\!\frac{2\Omega_{R}}{\omega}\!\!\right)\!\!\right)\!\!\right],  \label{fbeta}
\end{equation}
where has $\partial_{\beta}F(\beta)>0$ with $x<1.5$  ( $x$ is $0.499$ and $0.509$ for $^{87}\rm{Rb}$ and $^{23}\rm{Na}$ BECs, respectively). Owing to  $\partial_{\beta}F(\beta)>0$ and $\beta\in\left[0,1/4\right]$, $F(1/4)$ is larger than $F(0)$.

The first and second derivative of per particle energy $\epsilon_{k_{1}}$ for $k_{1}$ are given as
\begin{eqnarray}
\epsilon^{\prime}_{k_{1}}&\!=\!&k_{1}\left(\frac{\Omega^{\ast}_{0}}{2|\alpha| k_{0}\sqrt{\alpha^{2}k^{2}_{0}-k^{2}_{1}}}-\frac{F(\beta)}{\alpha^{2}k^{2}_{0}}\right),\label{par-k1} \\
\epsilon^{\prime\prime}_{k_{1}}&\!=\!&\!\frac{\epsilon^{\prime}_{k_{1}}}{k_{1}}
+\!\frac{\Omega^{\ast}_{0}k^{2}_{1}}{2|\alpha| k_{0}\!\left(\alpha^{2}k^{2}_{0}\!-\!k^{2}_{1}\right)^{\frac{3}{2}}}.\label{par-k12}
\end{eqnarray}
With the help of Eqs.~(\ref{par-k1}), and (\ref{par-k12}),  we demonstrate that there are three cases, i.e., $F(1/4)<0$, $F(0)<0<F(1/4)$, and $ F(0)>0$.

In the case $F(1/4)<0$, the energy minima will occur at $k_{1}=0$, owing to $\epsilon^{\prime}_{k_{1}}>0$. Thus, the per particle  energy $\epsilon$ with $k_{1}=0$ is
\begin{equation}
\epsilon(k_{1}=0,\beta)=\frac{1}{2}-\frac{\Omega^{\ast}_{0}}{2}+ \left(G_{1}-G_{2}\xi\right)\left(1+2\beta\right),\label{e-5}
\end{equation}
where has
\begin{equation}
G_{1}-G_{2}\xi =\frac{a}{32}\left(4+8x\right).
\end{equation}
 Due to $\left(G_{1}-G_{2}\xi\right)>0$ for the system with positive $x$, the minimum of Eq.~(\ref{e-5}) will be at $\beta=0$. In such case, the ground state is zero momentum phase (phase \Rmnum{3}), and the corresponding  per particle  energy is given by
\begin{equation}
\epsilon(k_{1}=0,\beta=0)=\frac{1}{2}-\frac{\Omega^{\ast}_{0}}{2}+ \left(G_{1}-G_{2}\xi\right).\label{e-4}
\end{equation}
In the region of $x<1.5$, $F(1/4)$ is always greater than $0$. Therefore, $F(1/4)<0$ can not be satisfied in spin-orbit coupled $^{87}$Rb or
$^{23}$Na BECs.

In the case $F(0)<0<F(1/4)$, systems only have two possible ground states, i.e., phase \Rmnum{1} and phase \Rmnum{3}.
 The reason why phase \Rmnum{2} disappears  will be expounded in the bellow. We assume that  $F(\beta^{\ast})=0$ is satisfied at the point $\beta^{\ast}$. From Eq.~(\ref{par-k1}), we find that extreme point of the per particle energy $\epsilon$ is $k_{1}=0$ when $\beta$ stays in the regions $\left(0,\beta^{\ast}\right]$, and extreme points are $k_{1}=0$, $k_{1}=\pm k^{\ast}$ when $\beta$ stays in the regions $\left(\beta^{\ast},1/4\right]$ with the condition of  $\Omega^{\ast}_{0}< 2F(\beta)$, where $k^{\ast}$ has the form
\begin{equation}
k^{\ast}=\alpha k_{0}\sqrt{1-\frac{\left(\Omega^{\ast}_{0}\right)^{2}}{4F^{2}(\beta)}}.
\end{equation}
With the help of the Eq.~(\ref{par-k12}), the minimum energy  occurs at  $k_{1}=k^{\ast}$ when $ k^{\ast}$ exist or at $k_{1}=0$ when $ k^{\ast}$ is inexistent. Next, we qualitatively analyze which phase will be the ground state at a given Raman frequency $\Omega^{\ast}_{0}$. At
$\Omega^{\ast}_{0}\!>\!2F(1/4)$, the ground state is phase \Rmnum{3}, owing to $\partial_{\beta}
\epsilon\left(k_{1}=0,\beta\right)> 0$. When $2F\left(1/4\right)> \Omega^{\ast}_{0}= 2F\left(\beta_{2}\right) >
2F\left(\beta^{\ast}\right)$  are satisfied, there are two cases. If the variational parameter $\beta^{\prime}\in\left[0,\beta_{2}\right]$ is
satisfied, the systems stay in states $\epsilon\left(k_{1}=0,\beta^{\prime}\right)$, then it is easy to know that the systems staying in phase \Rmnum{3}.  If the variational parameter $\beta^{\prime}\in\left(\beta_{2},1/4 \right]$ is satisfied, the per particle energy of systems is $\epsilon\left(k_{1}= k^{\ast},\beta^{\prime}\right)$. For $\beta^{\prime}$ in the region$\left(\beta_{2},1/4 \right]$,  in order to analyze the ground state, we need to introduce the first and second derivation of $\epsilon\left(k_{1}= k^{\ast},\beta^{\prime}\right)$, which are
\begin{eqnarray}
\partial_{\beta}\epsilon\left(k_{1}=
k^{\ast},\beta\right)&=&\frac{\left(\Omega^{\ast}_{0}\right)^{2}}{8F^{2}(\beta)}\left[\partial_{\beta}F(\beta)\right]
-4G_{2},\label{par-1}\\
\partial^{2}_{\beta}\epsilon\left(k_{1}= k^{\ast},\beta\right)
&=&-\frac{\left(\Omega^{\ast}_{0}\right)^{2}}{4F^{3}(\beta)}\left[\partial_{\beta}F(\beta)\right]^{2}<0.\label{par-2}
\end{eqnarray}
With the help of Eq.~(\ref{par-1}), it is easy to know that the systems only stay in the trivial states (phases \Rmnum{2} or \Rmnum{3}) for $G_{2} <0$. Here, we are only interested in the case of $G_{2} >0$. Due to $\partial^{2}_{\beta}\epsilon\left(k_{1}= k^{\ast},\beta\right)<0$, the minimum is achieved at the endpoint, i.e., $\beta^\prime=1/4$, and then the systems stay in phase \Rmnum{1}.  Due to the above-mentioned arguments, we know that the systems have two possible ground state either phases \Rmnum{3} or phase \Rmnum{1} with the condition $2F\left(1/4\right)> \Omega^{\ast}_{0}= 2F\left(\beta_{2}\right) > 2F\left(\beta^{\ast}\right)$. In conclusion, the possible ground state is  either phase \Rmnum{1} or phase \Rmnum{3} at the condition of
$F(0)<0<F(1/4)$.

In the case $F(0)>0$, the systems will have three possible  phases \Rmnum{1}-\Rmnum{3}.  The argument for this case is similar to the case of  $ F(0)<0 < F(1/4)$. With simple argument, we find that the possible ground state is phase \Rmnum{1} or \Rmnum{2}, phase
\Rmnum{1} or \Rmnum{3}, and phase  \Rmnum{3} with the Raman frequency satisfying the condition of $\Omega^{\ast}_{0} <  2F(0)$,
$2F(0)< \Omega^{\ast}_{0} <  2F(1/4)$, and $\Omega^{\ast}_{0} > 2F(1/4)$, respectively.
\begin{figure}[b]
\centering
\includegraphics[width=0.90\linewidth]{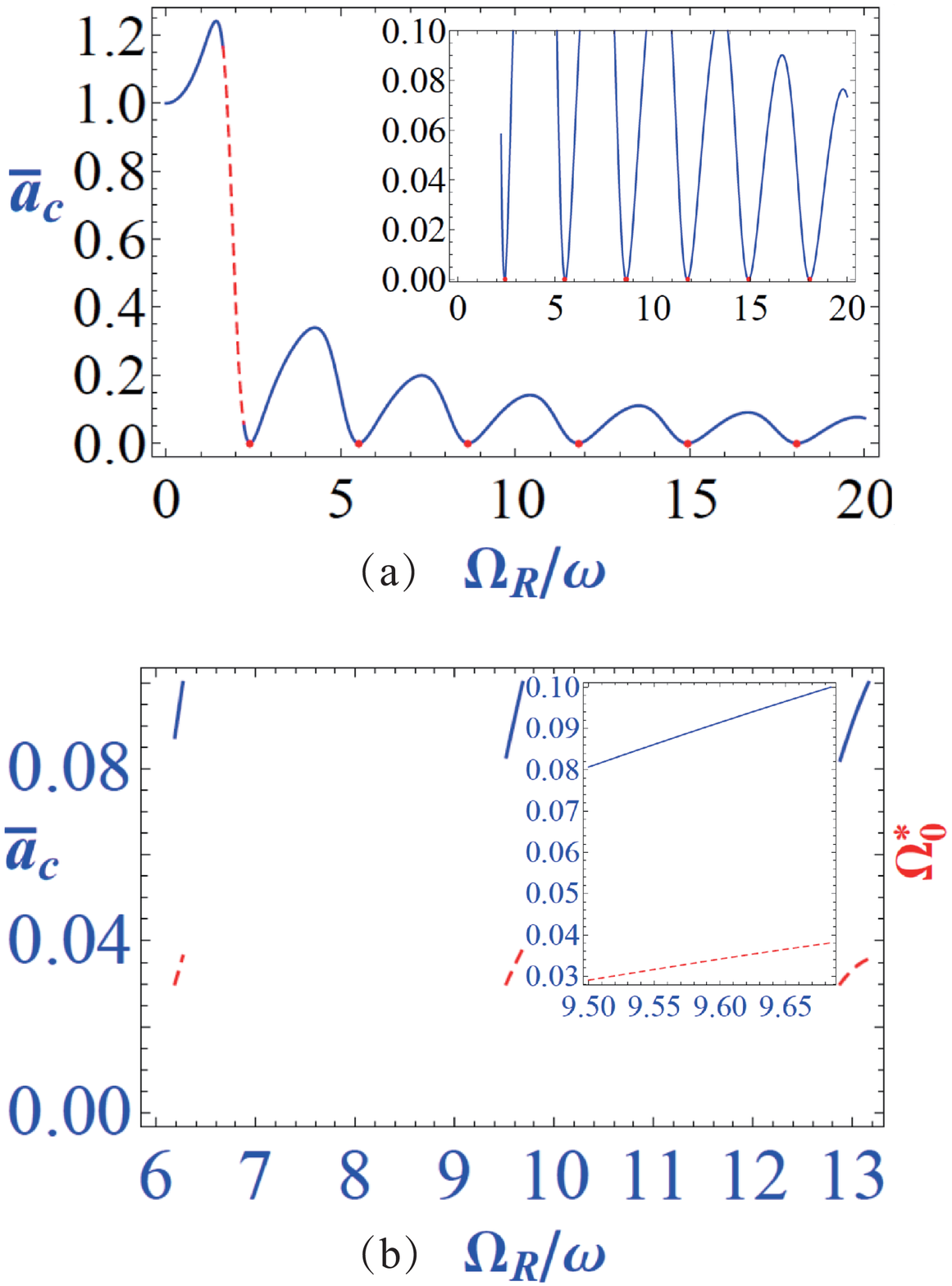}
\caption{(a) The tricritical value $\bar{a}_{c}=a_{c}/a^{0}_{c}$ for $^{87}\rm{Rb}$ atoms changes with dimensionless Raman coupling strength  $\Omega_{R}/\omega$. The red dashed line indicate the case $F(\beta=0,\bar{a}_{c})<0$ [where phase \Rmnum{2} does not exist (see the case F(0) < 0 < F(1/4) of the Section~\ref{sec-32}], thus this critical density is not real critical density. The red dots indicate no SOC and inserting picture is a function $\bar{a}_{c}$ with the constraint of $\bar{a}_{c}<0.1$. Here $a^{0}_{c}\approx 860$ is the quantum tricritical point value for undriven systems and the corresponding tricritical density is about $10^{17}\rm{cm}^{-3}$. (b) We restrict the quantum tricritical points within the region of $(\Omega^{\ast}_{c}>0.03,\bar{a}_{c}< 0.1 )$ and show $\bar{a}_{c}$ (blue lines), $\Omega^{\ast}_{c}$ (red dashed lines) as a function with $\Omega_{R}/\omega$, where the insert is and enlarged figure of the middle part figure.}
\label{critical3}
\end{figure}

\begin{figure*}[t]
\centering
\includegraphics[width=0.92\linewidth]{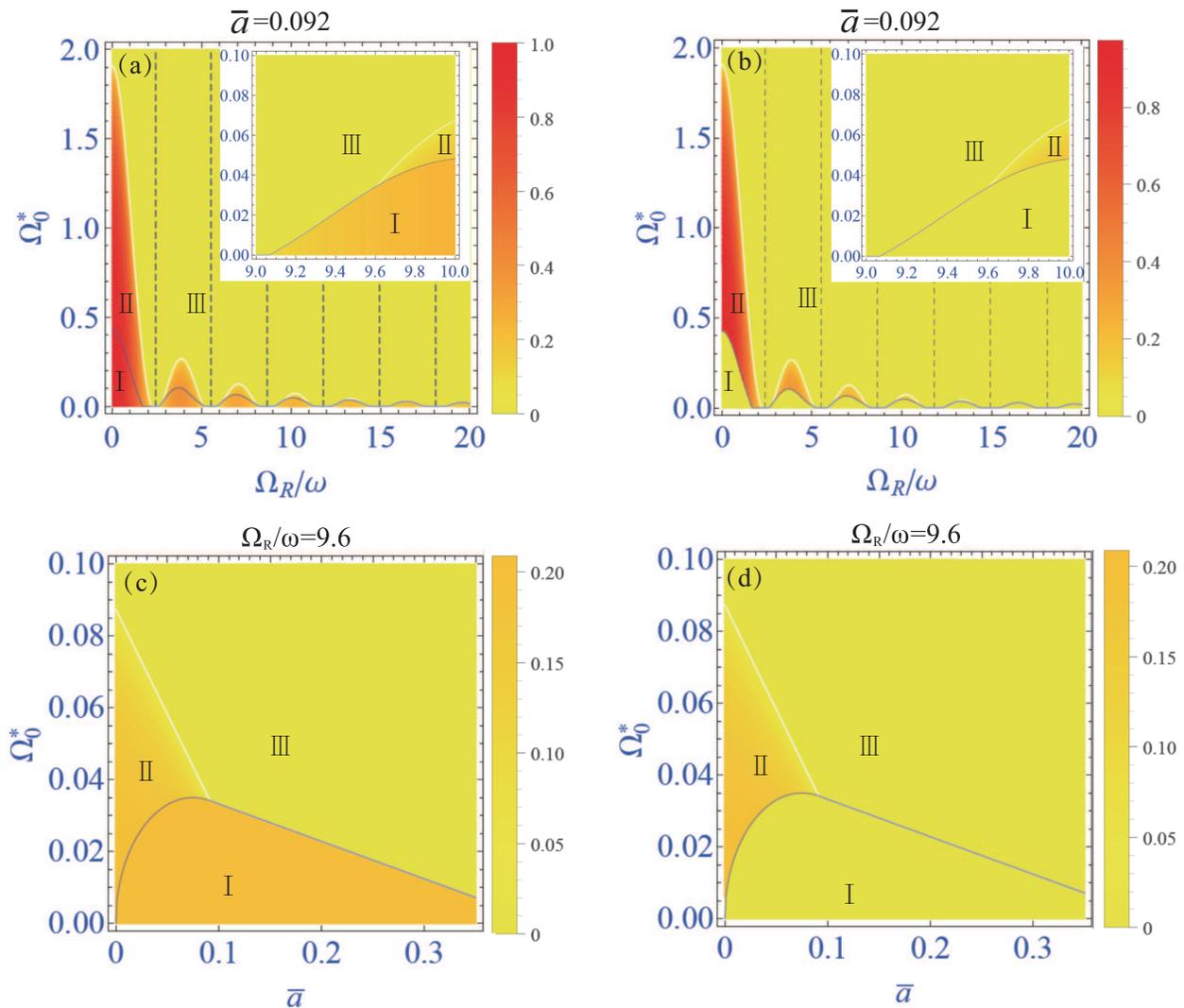}
\caption{(a)[(c)] $|k_{1}|/k_{0}$  as a functions with $\Omega^{\ast}_{0}$ and $\Omega_{R}/\omega$ [$\bar{a}$ (divided by $a_{c}^{0}$)] for fixed $\bar{a}$ ($\Omega_{R}/\omega$).  (b)[(d)] spin polarization $|\langle \sigma_{z}\rangle|$ as a functions with $\Omega^{\ast}_{0}$ and
$\Omega_{R}/\omega$ ($\bar{a}$) for fixed $\bar{a}$ ($\Omega_{R}/\omega$). The inserts are the amplification  of the diagrams at region of $9\leq\Omega_{R}/\omega\leq 10$.  The vertical  black dashed lines indicate the  amplitude of spin orbit coupling is zero, and then the systems stay in phase \Rmnum{3}. In (c) and (d), the $\Omega^{\ast}_{0}$ is the linear function of $\bar{a}$, and gradient is negative with large density. The phase transitions (\Rmnum{1}-\Rmnum{2} and \Rmnum{1}-\Rmnum{3} ) are discontinuous, while phase transition (\Rmnum{2}-\Rmnum{3}) is continuous. }
\label{rb3}
\end{figure*}
With the above-mentioned qualitative arguments, it is easy to know that there are three possible phases, i.e., phases  \Rmnum{1}, \Rmnum{2} and \Rmnum{3} as the ground state in tunable spin-orbit coupled BECs.  After the qualitative discussion, the quantitative phase
boundaries can be obtained by comparing the energy of phases \Rmnum{1}, \Rmnum{2} and \Rmnum{3}. In the most interesting case $G_{2}>0$, the systems will be in the phase \Rmnum{1} for small values of Raman coupling  strength $\Omega^{\ast}_{0}$. Under the condition
\begin{equation}
\alpha^{2}> G_{2}\left(4+2\xi\right)+\frac{4\left(G_{2}\right)^{2}}{G_{1}-G_{2}\xi}, \label{con1}
\end{equation}
the systems will undergo  phase transition from \Rmnum{1} to \Rmnum{2} at the Raman coupling
\begin{eqnarray}
\Omega^{\ast,\rm{\Rmnum{1}-\Rmnum{2}}}_{0}
&=&2\left[\frac{2G_{2}}{G_{1}+G_{2}\left(2-\xi\right)}\left(\alpha^{2}-2G_{2}\left(1+\xi\right)\right)\right. \nonumber \\
&&\times \left(\alpha^{2}+G_{1}-3G_{2}\xi\right)\bigg]^{\frac{1}{2}}.\label{ek1}
\end{eqnarray}
Increasing $\Omega^{\ast}_{0}$, the systems will remain in phase \Rmnum{2}, until at the Raman frequency
\begin{equation}
\Omega^{\ast,\rm{\Rmnum{2}-\Rmnum{3}}}_{0}
=2\left(\alpha^{2}-2G_{2}\left(1+\xi\right)\right).\label{ek2}
\end{equation}
If condition Eq.~(\ref{con1}) is not satisfied, the phase \rm{\Rmnum{2}} will disappear. The systems will directly enter phase \rm{\Rmnum{3}} from
\rm{\Rmnum{1}} at frequency
\begin{eqnarray}
\Omega^{\ast,\rm{\Rmnum{1}-\Rmnum{3}}}_{0}&=&2\left(\alpha^{2}+G_{1}-3G_{2}\xi\right)
-2\left[\left(\alpha^{2}+G_{1}-3G_{2}\xi\right)\right.
 \nonumber \\
&& \left.\times \left(G_{1}-G_{2}\xi\right)\right]^{\frac{1}{2}}.\label{ek3}
\end{eqnarray}

In the strong coupling (or high density) limit $G_{1}\gg \alpha^{2}$, the asymptotic behavior of Eq.~(\ref{ek3}) is $\alpha^{2}-2G_{2}\xi$  is not a constant value that is different form the undriven system \cite{li-1}. Therefore, in the strong coupling limit, the  $\Omega^{\ast,\rm{\Rmnum{1}-\Rmnum{3}}}_{0}$  is the linear  function of $a$ (or density)  and the gradient is negative with the fixed Raman coupling
strength $\Omega_{R}/\omega$.  At the below, by choosing the typical alkali BECs as the examples, i.e., $^{87}\rm{Rb}$ BECs and $^{23}\rm{Na}$ BECs,  the corresponding phase diagrams can be drawn via using the above-mentioned phase boundary Eqs.~(\ref{ek1}), (\ref{ek2}), and (\ref{ek3}).

\section{Application to tunable spin-orbit coupled $^{87}\rm{Rb}$ and $^{23}\rm{Na}$ BECs \label{sec-4}}
With the mean-field approach introduced in section~\ref{sec-3}, we now apply this framework to analytically investigate the phase diagram of tunable spin-orbit coupled $^{87}\rm{Rb}$ and $^{23}\rm{Na}$ BECs, respectively.
\subsection{phase diagrams of tunable spin-orbit coupled $^{87}\rm{Rb}$ BECs \label{sec-41}}
Before presenting the phase diagram of tunable spin-orbit coupled $^{87}\rm{Rb}$ BECs, we introduce the quantum tricritical point $a_{c}$ [obtained by taking the equal sign in Neq.~(\ref{con1}), where the phase (II) disappears] as a function of dimensionless Raman coupling strength  $\Omega_{R}/\omega$. In order to compare with the undriven systems, we use the dimensionless $\bar{a}_{c}=a_{c}/a^{0}_{c}$ instead of $a_{c}$. Here $a^{0}_{c}\approx 860$ is the critical value for undriven spin-orbit coupled $^{87}\rm{Rb}$ BECs. The corresponding critical density of $a^{0}_{c}\approx 860$ is about $10^{17}\rm{cm}^{-3}$, which is four orders of magnitude larger than the typical density \cite{density}. In Fig.~\ref{critical3}(a),  the critical value $\bar{a}_{c}$ as a function of  $\Omega_{R}/\omega$ is presented. By considering the difficulty of the implementation in experiment ( In experiment, $\Omega^{\ast}_{c}$ is not very small and $\bar{a}_{c}$ is not very large \cite{tunable-so}), we restrict the quantum tricritical points in the region $(\Omega^{\ast}_{c}>0.03, \bar{a}_{c}< 0.1 )$, tricritical Raman frequency $\Omega^{\ast}_{c}$ and the tricritical dimensionless $\bar{a}_{c}$ as a function of  $\Omega_{R}/\omega$ are shown in Fig.~\ref{critical3}(b). In our presented parameter region, the contrast \cite{Li} $(n_{\rm{max}}-n_{\rm{min}})/(n_{\rm{max}}+n_{\rm{min}})$ is only about $6\sim7 \times 10^{-4}$ in phase \Rmnum{1}. In order to directly observe these quantum tricritical points, experimenters need to enhance the measurement accuracy and to increase the densities of ultracold atoms.

In tunable spin-orbit coupled $^{87}\rm{Rb}$ BECs,  the phase diagrams with fixed density $\bar{a}=a/a_{c}^{0}=0.092$ or dimensionless Raman coupling strength  $\Omega_{R}/\omega=9.6$ are presented in Figs.~(\ref{rb3}). In these parameter regions, the critical density and Raman frequency $\Omega^{\ast}_{c}$ are not very small, therefore the quantum tricritical point will possibly be observed in future experiments. In Figs.~(\ref{rb3}), spin polarization $|\langle \sigma_{z}\rangle|$  and $|k_{1}|/k_{0}$  as functions of Raman frequency $\Omega^{\ast}_{0}$ and dimensionless Raman coupling strength  $\Omega_{R}/\omega$ ($\bar{a}$) in three different phases with given $\bar{a}=0.092$ ($\Omega_{R}/\omega=9.6$) are shown.
The spin polarization of $z$ direction can be calculated by
\begin{equation}
|\langle \sigma_{z}\rangle|=\frac{|k_{1}|}{k_{0}}\Big||C_{1}|^{2}-|C_{2}|^{2}\Big|.
\end{equation}
For fixed density [see Figs.~\ref{rb3}(a), and (b)], the parameter regions of the phase \Rmnum{1} and phase \Rmnum{2} are quasi-periodically  shrunken with increasing dimensionless Raman coupling strength  $\Omega_{R}/\omega$. Moreover, a series of  quantum tricritical points emerge
quasi-periodically with increasing $\Omega_{R}/\omega$. When the density is lager critical density (density at quantum tricritical point), the $\Omega^{\ast}_{0}$ is the linear function of $\bar{a}$ with negative slope [see Figs.~\ref{rb3} (c), and (d)]. This linear function feature is different form the undriven systems. In addition, in tunable spin-orbit coupled $^{87}\rm{Rb}$, the transitions (\Rmnum{1}-\Rmnum{2} and \Rmnum{1}-\Rmnum{3} ) are discontinuous and transition (\Rmnum{2}-\Rmnum{3}) is a continuous phase transition. It is in good agreement with transition types between these phases in undriven spin-orbit coupled $^{87}\rm{Rb}$.
\begin{figure}[b]
\centering
\includegraphics[width=1.0\linewidth]{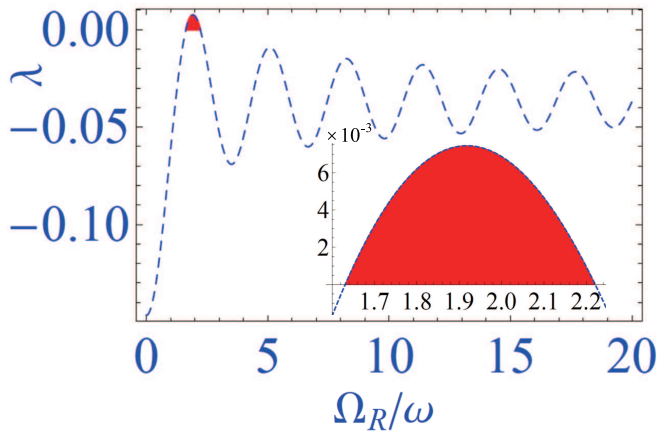}
\caption{The miscibility of the two spin components of $^{23}\rm{Na}$ atom. The $\lambda$ as a function with the $\Omega_{R}/\omega$ are presented. Although the system is immiscible for undriven case, the two components of $^{23}\rm{Na}$ atom are mixture stable in the region of red color.  The inset is an enlarged figure of $\lambda$.}
\label{stable}
\end{figure}
\begin{figure*}[t]
\centering
\includegraphics[width=0.9\linewidth]{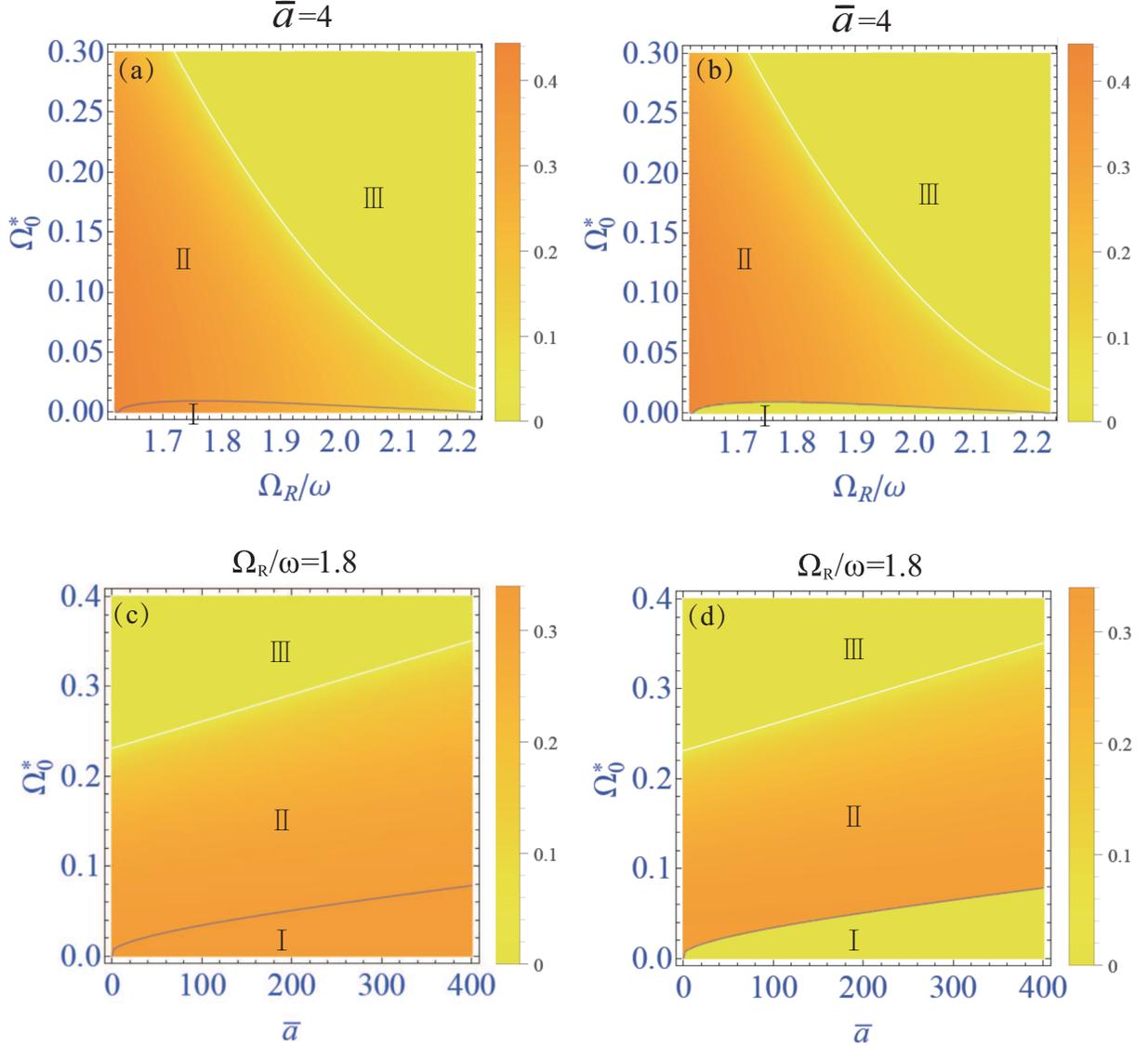}
\caption{$|k_{1}|/k_{0}$ [(a) and (c)]  and spin polarization $|\langle \sigma_{z}\rangle|$ [(b) and (d)] as  a function with $\Omega^{\ast}_{0}$, $\bar{a}$ and $\Omega_{R}/\omega$.  In (a) and (b), the phase \Rmnum{1} only have small region with small $\bar{a}$.  In (c) and (d), the  $\Omega^{\ast}_{0}$ at boundary of phase \Rmnum{1}-\Rmnum{2} is the linear  function of $\bar{a}$  and gradient is positive with large density. }
\label{rb4}
\end{figure*}
\subsection{phase diagrams of tunable spin-orbit coupled $^{23}\rm{Na}$ BECs  \label{sec-42}}
In tunable spin-orbit coupled $^{23}\rm{Na}$ BECs, the  phase diagrams can also be obtained by a similar method.  The state $\!\left|\! \right. F=1, m_{F}=0\rangle$ ( $\!\left|\! \right.F=1, m_{F}=-1\rangle$) of $^{23}\rm{Na}$ atom can be mapped to the pseudo-spin-up state $\!\left|\! \right. \uparrow\rangle$ ( pseudo-spin-down $\!\left|\!\right.\downarrow\rangle$). The scattering lengths for different spins are presented as \cite{spinor} $a_{\uparrow\uparrow}= C_{0}$ and $a_{\downarrow,\downarrow}=a_{\uparrow,\downarrow}= C_{0}+C_{2}$, where have \cite{spinor1} $C_{0}=(2a_{2}+a_{0})/3$, $C_{2}=(a_{2}-a_{0})/3$, $a_{2}=(52.98 \pm 0.40) a_B$ and $a_{0}=(47.36 \pm 0.80)a_B$ ($a_B = 0.0529 \rm{nm}$ is Bohr radius). Using above parameters, we can obtain $x\approx 0.509$.

Before calculating the phase diagrams of the tunable spin-orbit coupled $^{23}\rm{Na}$ BECs, we will talk about the miscibility of two components of $^{23}\rm{Na}$ atoms. We know that if $\lambda=g^{\ast}_{\uparrow,\uparrow}g^{\ast}_{\downarrow,\downarrow}-(g^{\ast}_{\uparrow,\downarrow})^{2}\!>\!0$ is satisfied, a
homogeneous mixture of two components is stable \cite{stable}. If we make the  naive  assumption that this criterion is also correct for driven systems, the systems are miscible in some parameter region [see Fig.~(\ref{stable})]. This is quite different from the undriven systems. However, there are new interaction terms
$-(g^{\ast}/2)(\psi^{\dag}_{\downarrow}\psi^{\dag}_{\downarrow}\psi_{\uparrow}\psi_{\uparrow}
+H.c.)$ in this effective Floquet Hamiltonian. The energy of this term
$(\psi^{\dag}_{\downarrow}\psi^{\dag}_{\downarrow}\psi_{\uparrow}\psi_{\uparrow} +H.c.)$ is always positive in three possible phases, moreover the energy of this term in stripe phase is smaller than that in the other phases. It is easy to know that $-g^{\ast}$ is positive for $^{23}\rm{Na}$ atoms and negative for $^{87}\rm{Rb}$  atoms. Therefore, this new interaction can lead to that $^{23}$\rm{Na} ($^{87}$\rm{Rb}) BECs prefer (dislike) to stay in stripe phase. In short, although the new interaction term  exists in tunable spin-orbit coupled $^{23}\rm{Na}$ BECs, $\lambda> 0$  can also be considered as a condition to estimate the miscibility of the two spin components for tunable spin-orbit coupled $^{23}\rm{Na}$ BECs.

In tunable spin-orbit coupled $^{23}\rm{Na}$ BECs,  the phase diagrams  with fixed density $\bar{a}=a/a^{0}=4$ \cite{na}  or dimensionless Raman coupling strength  $\Omega_{R}/\omega=1.8$ are presented in Figs.~(\ref{rb4}). With fixed $\bar{a}=4$ ($\Omega_{R}/\omega=1.8$), spin polarization $|\langle \sigma_{z}\rangle|$  and $|k_{1}|/k_{0}$  as a function of Raman frequency $\Omega^{\ast}_{0}$ and dimensionless Raman coupling strength $\Omega_{R}/\omega$ (density $\bar{a}$ ) in three different phases  are shown in Figs.~(\ref{rb4}). The transition (\Rmnum{1}-\Rmnum{2}) is discontinuous and transition (\Rmnum{2}-\Rmnum{3}) is a continuous phase transition. If the density [see Figs.~\ref{rb4}(a), and (b)] is fixed at small value, e.g., $\bar{a}=a/a^{0}=4$, the region of phase \Rmnum{1} is very small at the current experimental density. When dimensionless Raman coupling strength is given such as $\Omega_{R}/\omega=1.8$ [see Figs.~\ref{rb4}(c), and (d)],  Neq.~(\ref{con1}) is always satisfied in the miscible parameter region [the region of red color in Fig.~(\ref{rb3})] that means the phase \Rmnum{2} is always there. Thus, quantum tricritical point will not appear in $^{23}\rm{Na}$ systems. Moreover, $\Omega^{\ast,\rm{\Rmnum{1}-\Rmnum{2}}}_{0}$ is a linear function of $\bar{a}$ (or density) with positive slope in the high density regions  [see Figs.~\ref{rb4}(c), and (d)].  This linear behavior of $\Omega^{\ast,\rm{\Rmnum{1}-\Rmnum{2}}}_{0}$  can be understood by the fact that  the asymptotic behavior of $\Omega^{\ast,\rm{\Rmnum{1}-\Rmnum{2}}}_{0}$ with high density is $\left[\alpha^{2}+ \frac{1}{2}\left(G_{1}-G_{2}(5\xi+2)\right)\right]\left[2G_{2}/\left(G_{1}+G_{2}(2-\xi)\right)\right]^{1/2}$ which is $\propto a$  for fixed $\Omega_{R}/\omega$.

\section{summary\label{sec-5}}
In conclusion, the effective Floquet Hamiltonian of tunable spin-orbit coupled BECs with two-body interactions has been demonstrated. And then, the phase boundaries of tunable spin-orbit coupled  BECs have been studied by using variational wave function to obtain the ground state of this effective Floquent Hamiltonian. By taking tunable spin-orbit coupled $^{87}\rm{Rb}$  and $^{23}\rm{Na}$ BECs as examples, the phase diagrams are also presented. In contrast with the undriven systems, the characteristic features of the phase diagrams of tunable spin-orbit coupled BECs are presented in the following.

In tunable spin-orbit coupled $^{87}\rm{Rb}$ BECs, the critical density $n_{c}$  can be reduced dramatically in some parameter region. Therefore, the prospect of observing this intriguing quantum tricritical point is optimistic in this tunable spin-orbit coupled $^{87}\rm{Rb}$ BECs.  At fixed density, the quantum tricritical points emerge quasi-periodically with increasing the dimensionless Raman coupling strength $\Omega_{R}/\omega$. Although the phase diagrams are similar to the undriven spin-orbit coupled $^{87}\rm{Rb}$ BECs, in the strong coupling limit $G_{1}\gg \alpha^{2}$, the asymptotic behaviour of $\Omega^{\ast,\rm{\Rmnum{1}-\Rmnum{3}}}_{0}$  is $\alpha^{2}-2G_{2}\xi$, which is not a constant value.

In tunable spin-orbit coupled $^{23}\rm{Na}$ BECs,  it is surprising that two hyperfine states of  $^{23}\rm{Na}$ atoms are miscible in some parameter regions. In these miscible regions,  $^{23}\rm{Na}$  systems can stay in the stripe phase with small Raman frequency and experimental level density. The regions of stripe phase can be expanded when the density is increased. In contrast to $^{87}\rm{Rb}$ systems, there is no quantum tricritical point in such $^{23}\rm{Na}$ systems. These characteristic features will be observed  with improving the measurement accuracy.

\section*{Acknowledgments}
 We would like to thank Yan Chen, Yun Li, Xia-Ji Liu and Hui Hu  for useful discussions. We wish also to thank
Dan Bo Zhang and Wan-Li Liu for reading and providing useful comments on this
manuscript. This work was supported by  the National Natural Science Foundation of China (Grant Nos. 11947102), the  PhD research Startup Foundation of Anhui University (Grant No. J01003310) and the Open Project of State Key Laboratory of Surface Physics in Fudan University (Grant No. KF$2018\_13$).


\begin{thebibliography}{99}
\bibitem{density}In Yun's work \cite{li-1}, we know the critical density at the level of $10^{15}\rm{cm}^{-3}$ with the recoil energy $k^{2}_{0}/M_{Rb}=2\pi\times80\rm{Hz}$, but the current experimental $k^{2}_{0}/M_{Rb}$ is about $2\pi\times 3.677\rm{k Hz}$. In current experiment, the density value is about $ 10^{13}\rm{cm^{-3}}$.
\bibitem{na}In undriven $^{23}\rm{Na}$ systems, we choose $a_{0}\approx 0.1095$ with typical density $n=5\times 10^{13}\rm{cm}^{-3}$ and  typical recoil energy $k^{2}_{0}/M_{Na}=2\pi\times 13.9014\rm{kHz}$ (or $k_{0}=5.623\rm{\mu m}^{-1}$), where recoil momentum $k_{0}=5.623\rm{\mu m}^{-1}$ is the same with the recoil momentum of $^{87}\rm{Rb}$ BECs.
\end{thebibliography}
\end{document}